\documentclass[a4paper]{article}
\usepackage{RR}
\RRNo{6251}
\usepackage{hyperref}
\usepackage{color}
\usepackage{amsmath}
\usepackage{newalg}
\usepackage{chemarrow}
\RRdate{Juin 2007}

\RRauthor{
Yann Busnel\thanks{IRISA / UR1 - ENS Cachan}%
  \and
Marin Bertier\thanks{IRISA / INSA Rennes}%
  \and
Eric Fleury\thanks{CITI / INSA Lyon}%
  \and
Anne-Marie Kermarrec\thanks{IRISA / INRIA Rennes}%
}
\authorhead{Y. Busnel, M. Bertier, E. Fleury and A.-M. Kermarrec}
\RRtitle{GCP: Mise \`a jour \'epid\'emique de logiciels pour les r\'eseaux de capteurs mobiles, larges \'echelles}
\RRetitle{GCP: Gossip-based Code Propagation for Large-scale Mobile Wireless Sensor Networks}
\titlehead{GCP: Gossip-based Code Propagation for mobile WSN}
\RRresume{GCP est un systeme de mise \`a jour de code automatique pour r\'eseaux de capteurs mobiles, utilisant le concept de diffusion \'epid\'emique d\'eve\-lop\-p\'es dans le cadre de r\'eseaux filaires. \\
Ce rapport pr\'esente la conception et l'\'evaluation de GCP ({\em Gossip-based Code Propagation}), protocole propos\'e dans le cadre de ces travaux de recherches. \\
\\
Celui-ci est \'evalu\'e par comparaison avec des algorithmes traditionnels de dissemination de donn\'ees. Les r\'esultats de simulations sont fond\'es \`a la fois sur des traces r\'eelles et g\'en\'er\'ees, permettant de montrer l'efficacit\'e de GCP tant dans la vitesse de propagation que dans l'\'equilibrage des charges sur le r\'eseau.
}
\RRabstract{Wireless sensor networks (WSN) have recently received an increasing interest. They are now expected to be deployed for long periods of time, thus requiring
software updates. Updating the software code automatically on a huge
number of sensors is a tremendous task, as ''by
hand'' updates can obviously not be considered, 
especially when all participating sensors are embedded on
mobile entities.\\
\\
In this paper, we investigate an approach to automatically update
 software in mobile sensor-based application when no localization
 mechanism is available. We leverage the peer-to-peer cooperation
 paradigm to achieve a good trade-off between reliability and
 scalability of code propagation. More specifically,
we present the design and evaluation of
 GCP ({\emph Gossip-based Code Propagation}), a distributed
 software update algorithm for mobile wireless sensor networks. GCP
 relies on two different mechanisms (piggy-backing and forwarding control)
  to improve significantly the load balance
 without sacrificing on the propagation speed. 
We compare GCP against
traditional dissemination approaches. Simulation results based on
 both synthetic and realistic workloads show that GCP achieves a good
 convergence speed while balancing the load evenly between sensors.}
\RRmotcle{r\'eseaux de capteurs, informatique mobile, large \'echelle, 
diffusion, mise \`a jour de code, syst\`emes pair-\`a-pair, simulation}
\RRkeyword{Wireless sensor network, mobile computing, large scale,
diffusion, software update, peer-to-peer algorithm, simulation}
\RRprojets{Asap et Ares}
\RRtheme{\THNum} 
\URRennes  
\begin{document}
\makeRR   

\section{Introduction}
\label{toc:intro}

Recently, compact devices, called micro-electro mechanical systems
(MEMS), have appeared. Such devices combine small size, low cost,
adaptability, low power consumption, large scale and
self-organization. Equipped with wireless communication capability,
such appliances (called {\it mote node} or {\it sensor}\footnote{We
will called these devices {\it sensor} in the remaining of the
paper}) together form a
wireless sensor network (WSN). Due to their tiny size, sensors possess
slim resources in term of memory, CPU, energy, {\emph
etc}.~\cite{Akyildiz2002A-Survey-on-Sen, Rentala2001Survey-on-senso}.

The increasing interest in WSNs is fundamentally due to their
reliability, accuracy, flexibility, cost effectiveness and ease of
deployment characteristics. Such WSNs can be deployed for monitoring purposes
for example. For example we can cite {\it Ecosystem
monitoring}, {\it Military} (battlefield surveillance, enemy tracking,
\ldots), {\it Biomedical and health monitoring} (cancer detector,
artificial retina, organ monitor, \ldots), {\it Home} (childhood
education, smart home/office environment, \ldots).

Sensors may be deployed both in static and dynamic environments.
They are usually deployed for a long period of time, during
which the software may require updates. While efficient solutions to
software update may be deployed in fixed WSNs, this is far more
complex when sensors are embedded on mobile entities such as
people. In this paper, we consider this latter setting, namely a WSN deployed
over a group of people.

Given the potentially large number of participating sensors in WSN and
their limited resources, it is crucial to use fully decentralized
solutions and to balance the load as evenly as possible between
participating sensors. In that context, we investigate the use of the
P2P communication paradigm which turns out to be a relevant candidate
in this context.
Considering similarities between these two systems,
 Section~\ref{toc:p2p} investigates the
relevance of adapting the P2P paradigm to mobile WSN. Using epidemic-based
dissemination, we introduce a greedy protocol ({\it GCP: Gossip-based
Code Propagation}) balancing the dissemination load without
increasing diffusion time. GCP relies on \emph{piggy-backing} to save up energy and \emph{forwarding control} to balance the load among the nodes.

Code propagation (or reprogramming service)
has a lot in common with broadcast and data
dissemination~\cite{Chen2005Survey-on-Peer-,Intanagonwiwat2003Directed-diffus}
with an additional main-constraint. In broadcast, each message sent
before a node arrival can be ignored by this node. In software update
protocols, each new node in the network has to be informed of the
existence of the software's latest version as soon as possible to be
operational.

This paper is organized as follows. Section~\ref{toc:p2p} presents the
P2P paradigm and more specifically the epidemic principle.
Section~\ref{toc:algo} introduces the GCP algorithm and alternative approaches.
We  compared GCP with classical approaches by simulation and
depict the  results in Section~\ref{toc:simu}. Finally,
Section~\ref{toc:relatedWorks} introduces a short state of art of
the different domains cited above before concluding in
Section~\ref{toc:conclu}.

\section{Applying P2P algorithms to mobile WSN}
\label{toc:p2p}

Classical  peer-to-peer systems are  composed  of millions
 of  Personal Computers connected together 
 by  a  wired network and as opposed to 
 sensor networks are not limited by node capacity,
communication range or system size (see Table~\ref{tab:diffP2PWSN}).
In this  section  we promote    the idea  that   sensor networks   and
peer-to-peer  systems  are similar enough so that P2P solutions can be seriously considered in the context of WSNs.

\begin{table*}
\begin{center}
{\small
\begin{tabular}{|c|c|c|}
\hline
& {\bf Peer to peer systems}    & {\bf Sensor networks}\\
\hline
\hline
\multicolumn{3}{|l|}{\sc Similarities}\\\hline
{\it System size}       & Millions                                              & Thousands \\\hline
{\it Dynamicity}        & Connection/disconnection              & Mobility \\
                                        & Failure                                               & Failure (low power)\\\hline
\hline
\multicolumn{3}{|l|}{\sc Differences}\\
\hline
{\it Resource}          & Plentiful                                             & Tiny \\\hline
{\it Potential neighbourhood} & Chosen among the whole system & Impose \\
\hline
{\it Connectivity}              & Persistent                                    & Temporary\\
\hline
\end{tabular}
}
\end{center}
\caption{Similarity and divergence between P2P systems and WSNs}
\label{tab:diffP2PWSN}
\end{table*}

\subsection{Peer-to-Peer vs. mobile WSN}

On one  hand, sensors    and  personal computers   have   incomparable
resources: PCs  have large resources   in terms of CPU, storage  while
sensors are very  limited;   sensors have strong   energy constraints,
limiting  their  capability  to   communicate.
  On the other hand,  due to the node's resources compared
to  the size of  the system, no  entity is able  to  manage the entire
network in both systems. Every application  designed for both networks
requires a strong cooperation  between entities to  be able  to manage
the  network  and to  take  advantage  of  the entire  network.   Peer to peer
solutions heavily rely on such collaboration. 

The peer to peer communication paradigm has been clearly identified as
a key to scalability in wired systems. 
In a P2P system, each node may act both as a
client and a server, and knows only few other nodes. Each node is
logically connected to a subset of participating nodes forming a
logical overlay over the physical network.  With this local
knowledge, the resource aggregation and the load\footnote{The load is
composed of forwarding messages, storing data,  \emph{etc}.} are evenly
balanced between all peers in the systems.  Central points of failures
disappear as well as associated performance
bottlenecks~\cite{Castro2003Proximity-neigh, Xu2004Data-Management}.

In  a sensor network,  a node is able to communicate
only with  a  subset of  the  network within its communication
range and has to opportunity to ``choose''its neighbours.
 In addition, in a  mobile WSN, the neighbourhood of 
 a node changes according to its  mobility pattern.  
With the exception of the sensor's
energy  constraint,  the other  differences  between this  two systems
which have a direct impact on the algorithm behaviour is the multicast
advantage of the wireless medium. When a  sensor node sends a message,
this  message   can be received  by  every  node  in  its   direct
neighbourhood  while in a wired network  a message is received only by
the nodes which are explicitly designated in the message.

If solutions designed for  a large scale wired   network can not   be
applied directly in a sensor network, the peer-to-peer paradigm is 
implicitly used in  sensor networks.

\subsection{Epidemic algorithms}

Epidemic or gossip-based communication is well-known to provide a
simple scalable efficient and reliable way to disseminate
information~\cite{KerMasGan03IEEEtpds}. Epidemic protocols are based
on continuous information exchange between nodes. Periodically, each
node in the system chooses randomly a node in its neighbourhood to
exchange information about itself or its neighbourhood.  One of the
key results is that in a random graph, if the node's {\it view} (set of
knowing nodes, {\it i.e.} the local view of the system) is constraint
to $c = O(\log(n))$, where $n$ is the size of the network, it assures
that every node in the system receive a broadcast message with a
probability of $e^{-e^{-c}}$.

Based on unstructured P2P overlay\footnote{The logical layer on top of
  the  physical   one is   not constrained  to  a define  structure.},
gossip-based protocols can be successfully applied in WSNs.
Recently, several approaches based on  gossip  have been proposed  in the context of  WSNs~\cite{Kyasanur2006Smart-Gossip:-A,Levis2004Trickle:-A-Self,Wang2005Reliable-Gossip,Wang2006Gappa:-Gossip-B}.

The objective of this paper is to adapt such an approach to achieve efficient and reliable software dissemination in mobile WSN.

\section[GCP : Control flooding for mobile WSNs]{GCP: Introducing control in flooding as a miracle drug to mobile WSNs}
\label{toc:algo}

In mobile wireless sensor networks, routing and broadcasting is a
challenging task due to the network dynamicity. To the best of our
knowledge,  existing approaches do not deal with diffusion persistence
(\emph{cf}. Section~\ref{toc:relatedWorks}).

In the following,  we  consider a distributed   system consisting of a
finite   set  of mobile   sensor nodes which  are   not aware of their
geographic location.  The network may not be connected at any time 
as  at a  time
$t$,  a node  can  only communicate with   nodes in  its communication
range.  However, we consider that over an application duration,  given 
that there are an
infinite number of paths between two nodes, the
network  is  eventually    connected.    In  order to  discover    its
neighbourhood,  each  node  periodically\footnote{This  period  is   a
  parameter of the   system} broadcasts locally  {\tt  Hello} messages,
called {\it beacon}.

\subsection{GCP design}
\label{toc:gcp-algo}

Flooding paradigm is a simple way to disseminate informations. Used commonly in the network area, it consists in forwarding to everyone known a new received information.  Rather than applying classical flooding algorithms having an ideal speed propagation at the price of a high energy consumption,  we use the
epidemic communication paradigm, proposed in the context of P2P
systems. To this end, GCP is inspired from the flooding paradigm 
enhanced through\emph{Piggy-Backing} and \emph{Forwarding Control}.

\paragraph{Piggy-Backing mechanism}
In order to avoid unnecessary software transmissions, nodes have to be aware of the  software versions hold by their neighbours. To this end, each node simply piggy-backs its own version number into beacon messages. 
\paragraph{Forwarding Control mechanism}
In order to balance the load among node and increase the overall lifetime 
of a system, each node sends its  current  software version a limited
number of times. To this end,  each node owns a given number
of {\em tokens}, which value is a system parameter. Sending a software update
is worth a token. When  a node has  spent all its tokens, it  is not
allowed  to send  this version of  the  software. 
This number of tokens is associated to each version. Upon receiving a new version of the software, the number of tokens is set to the default initial value.
This mimics the behaviour of an epidemic protocol, where each node sends 
a predefined number of time a message (typically $\log(N)$,   $N$  being the size of the system)~\cite{KerMasGan03IEEEtpds}. Likewise, the default value of the number of tokens can
be set according the order of magnitude of WSN size.

\subsection{GCP algorithm}

\begin{figure}
\begin{center}
\includegraphics[width=0.5\textwidth]{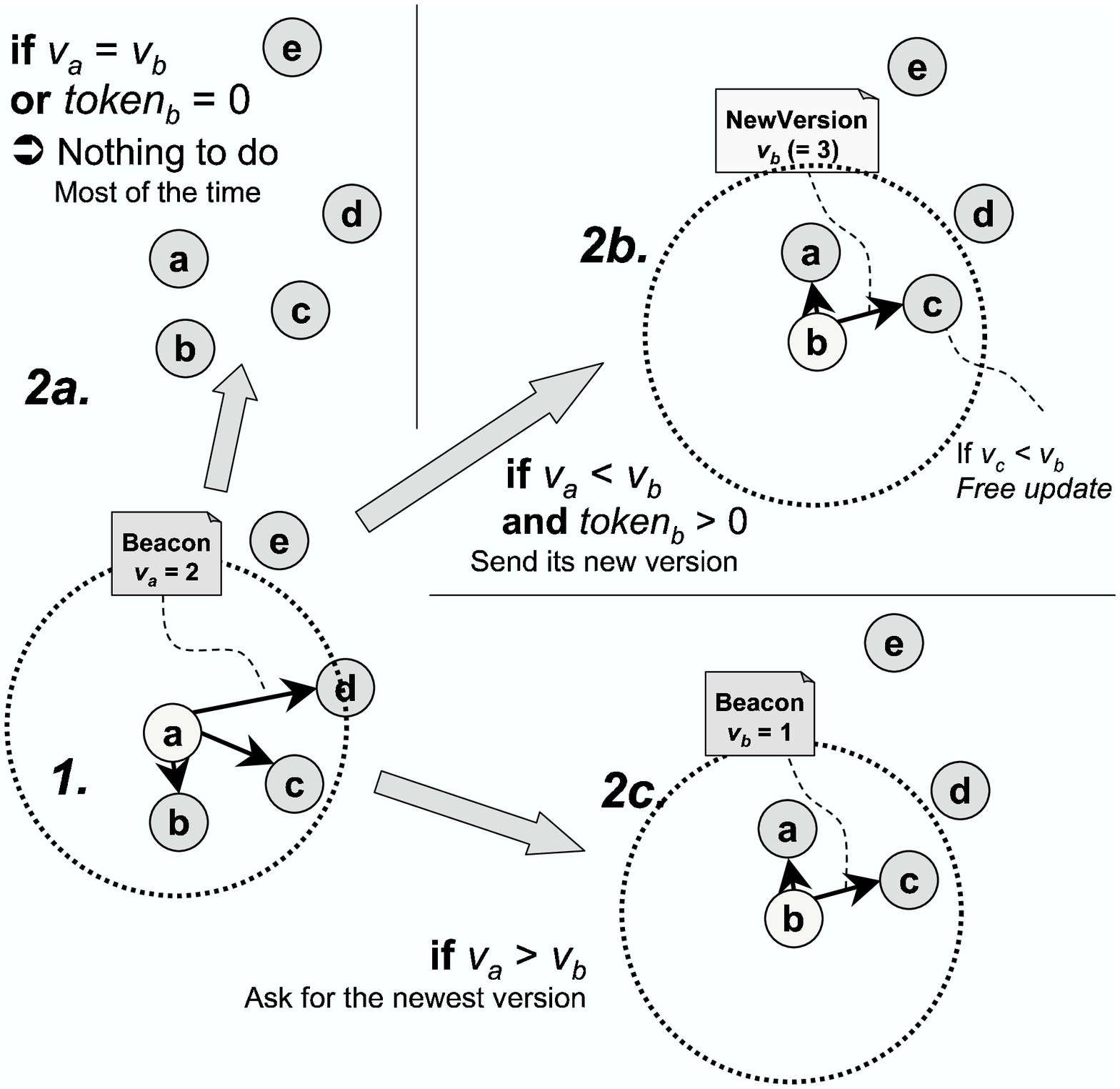}
\end{center}
\caption{Summarized behaviour of GCP}
\label{fig:gcpalgo}
\end{figure}

Figure~\ref{fig:gcpalgo} represents the three possible different cases
and the GCP behaviour. 

 Each node in the transmission range of  $a$ receives the beacon (In Figure~\ref{fig:gcpalgo}: nodes $b$,
$c$  and $d$; node $e$  is out of range).   A beacon message
received by  node $b$ is processed as follows:
\begin{enumerate}
\item[2a.] If $b$  owns the same version  as $a$ ($v_a=v_b$) due to the piggy-backing mechanism or $token_b = 0$ due to the forwarding control mechanism,  then no action is required. 
\item[2b.]  If $b$ owns a  more recent version than $a$  ($v_a<v_b$) due to the piggy-backing mechanism,
  and, if $b$ still holds some tokens ($token_b>0$) due to the forwarding control mechanism, it sends its
  own version to  $a$ thus consuming  a  token ($token_b$-{}-).  
Note that if other nodes, within the transmission range,  holding
 an older  version than $b$'s, they leverage the software update and update their own version (``free update'').
\item[2c.]  If $a$ owns a version newer than $b$ ($v_a>v_b$) due to the piggy-backing mechanism, the node
  $b$ sends   immediately  a beacon   message in  order  to request  a
  software update from $a$ while $a$ is still in its radio range. 
\end{enumerate}

\subsection{Alternative algorithms}

In order to assess the efficiency of GCP, we compare it against three other protocols, directly derived from wired networks. We briefly present those protocols in this section.

\begin{description}
\item{\bf The Flooding Protocol (FP)} Each time a node receives a beacon from another one, it sends its own version of the software, whether the node needs it or not. This  algorithm obviously leads to load unbalance and does not take into account energy consumption.   This
algorithm is presented here  because  it provides good  software
propagation speed. 
\item{\bf The Forwarding Control Protocol (FCP)}
This algorithm  is an enhancement of the  flooding protocol,
using the forwarding control mechanism.
\item{\bf The Piggy-Backing Protocol (PBP)}
This  last  algorithm  is an  enhancement  of  the flooding algorithm,
with the piggy-backing mechanism.
\end{description}

\subsection{Protocols in Pseudo-Code}

Thereafter, we introduce a formalized version of these protocols.

\begin{itemize}
\item $version$ and $version_r$ represent respectively the local and the remote version number of the software;
\item $software$ and $software_r$ represent respectively the local and the remote binary of the software;
\item $token$ represents the remaining number of tokens available on the local node;
\item $initialNumberOfTokens$ represents the initial number of tokens available on the local node.
\end{itemize}

FP is presented in Figure~\ref{fig:flooding}.

\begin{figure}[h]
\begin{center}
\fbox{\begin{minipage}{0.45\textwidth}
\begin{algorithm}{ReceiveBeacon}{\NIL}                  \algkey{send} (software, version)
\end{algorithm}

\begin{algorithm}{ReceiveSoftware}{software_r, version_r}       \begin{IF}{version_r > version}\\
                \algkey{updateCode}(software_r)\\
                version \= version_r
        \end{IF}\end{algorithm}
\end{minipage}}\end{center}
\caption{Flooding algorithm}
\label{fig:flooding}
\end{figure}

The GCP algorithm is presented in Figure~\ref{fig:greedy}. PBP has the same pseudo code without lines 2, 5 in {\sc ReceiveBeacon} and without line 5 in {\sc ReceiveSoftware}. FCP has the same pseudo code without lines 1, 6, 7 and 8 in {\sc ReceiveBeacon}.

Each time a node receives a beacon message, it compares its own
version number with the remote one. If it possesses a newer version,
it sends its code to its neighbourhood. Otherwise, if it owns an older
version, it sends immediately a beacon, to pull the newer version from
the remote node.

When a node receives a software version, it checks this version
number. If this version is a newer one, it replaces its own version by
the new one and reinitializes its token number to be able to forward it.

\begin{figure}[h]
\begin{center}
\fbox{\begin{minipage}{0.45\textwidth}
\begin{algorithm}{ReceiveBeacon}{version_r}   
\begin{IF}{version_r < version\\ 
\algkey{ and } token > 0}\\
                \algkey{send}(software, version)\\
                token \= token -1
   \ELSE \begin{IF}{version_r > version}\\
                \algkey{send} {\rm Beacon}(version)
           \end{IF}   \end{IF}\end{algorithm}

\begin{algorithm}{ReceiveSoftware}{software_r, version_r}       \begin{IF}{version_r > version}\\
                \algkey{updateCode}(software_r)\\
                version \= version_r\\
                token \= initialNumberOfTokens
        \end{IF}\end{algorithm}
\end{minipage}}
\end{center}
\caption{Gossip-based Code Propagation algorithm}
\label{fig:greedy}
\end{figure}

\subsection{Theoretical analysis}
\label{toc:theory}
We  compare these protocols along two  metrics: the software
propagation  speed and the  number of  versions  sent by  each  node.
This last metric reflects the load balance in the system.
 Some notations  are needed  and listed in
the following paragraph:
\begin{itemize}
\item $n_v$ represents the number  of upgrades during the WSN lifetime
  (how many times the software need to be updated);
\item $t$ represents the number of tokens available on each node;
\item  $n_{nh}$ represents the  average  size of a node  neighbourhood;
\item $d$ represents the duration of the experimentation;
\item $p_b$ represents the period of beacon emission;
\item $n_s$ represents the network size in terms of number of nodes;
\end{itemize}

For all of the following theoretical results, each equation presents the
upper bound average values.  To obtain more  precise results,  we may take
into account the diameter size, the topology and mobility model of the
WSN considered.

\paragraph{Load balancing}
For the load balancing, the analysis extracts the quantity of large message sent by each node, as software binary sending here. The following equations present the average number of software sent by one node during the whole deployment
of the WSN for each algorithm:
\begin{eqnarray}
  \label{eq:load}
  \textrm{Flooding algorithm:} &\sim& \frac{d}{p_b}\times n_{nh}\label{eqn:lFCP}\\
  \textrm{FCP algorithm:}   (lb_{FCP}) &\sim& (n_v + 1)\times t\label{eqn:lbFCP}\\
  \textrm{PBP algorithm:}   (lb_{PBP}) &\sim& n_v\times (ns - 1)\label{eqn:lbPBP}\\
  \textrm{GCP algorithm:}   (lb_{GCP}) &\sim& n_v\times t\label{eqn:lbgcp}
\end{eqnarray}


In the  flooding algorithm, at each  beacon reception,  a node sends its
own version. Equation (\ref{eqn:lFCP}) illustrates  that the number  of
time a node  sends its version  is equivalent to  the number of beacon
sent  by  each node multiplied  by the average number   of nodes in the
neighbourhood.

Both the FCP and GCP algorithms have  a bound on  the
quantity of time a  node can send  its software according to $n_v$ and
$t$.    For FCP,   as    the version  number  is    not communicated to
neighbourhood,  the first version  is sent as  a new one by each node.
Moreover, in the FCP algorithm,  tokens may be spent unnecessarily,
 as opposed to
GCP consuming only necessary tokens thanks  to the version number
contained in beacon messages. We assume that $n_s \gg t$
to optimize balancing for token rule using algorithms in a large scale
environment.

We can conclude that in the worst configuration:
$$lb_{FP} \gg lb_{PBP} \gg lb_{FCP} > lb_{GCP}$$
and in most cases:
$$lb_{FP} \gg lb_{PBP} > lb_{FCP} > lb_{GCP}$$

\paragraph{Propagation speed}
Considering the propagation speed, the analysis extracts the propagation of
a new software in the WSN. We consider the software update propagation by dividing
time into software forwarding step.

We start by comparing Flooding with GCP and PBP. Let $t$  be the current
step of propagation. Consider that at this step, the network is in the
same state (same nodes own the newest version of the protocol, same
positions of nodes and future moving,  \emph{etc}.) Obviously, using
flooding software is equivalent to forward the software to all nodes,
regardless of their need for it. 
At the opposite, in GCP and PBP,  a node sends the
newest software version only to nodes needing it. They exhibit a
similar propagation speed. GCP and PBP however provides software updates
``for free''. 
Effectively, if a node is located in the transmission range of the newest software, it will
update its software without requesting it explicitly. 
 By
recurrence on $t$, Flooding algorithm propagation speed ($ps_{FP}$) is greater
or equal to GCP and PBP ones (respectively $ps_{GCP}$ and $ps_{PBP}$).

Considering PBP and GCP, they are equivalent in the most common case.
But, PBP may have slightly better speed propagation in case of the
meeting set is unbalanced (one node is transmit the newest version
most of the time  \emph{i.e.} this node is consuming a large part of
its power). The GCP propagation speed can be slowed down as this
potential node does not have enough token. However, this case does not
balance the load among the network with PBP, which is one of our
objective.

We are now considering GCP and FCP algorithms. By choosing an appropriate
initial number of tokens, the propagation speed of these two
algorithms can approach the ideal one. Furthermore, most of the time,
using FCP algorithm could waste a non-negligible number of tokens by
sending unnecessarily the software.
 The FCP algorithm propagation speed
($ps_{FCP}$) can often be strictly lower of the GCP one.

So, we can conclude that in the worst configuration:
$$sp_{FP} \geq sp_{PBP} \geq sp_{GCP} \geq sp_{FCP}$$
and most of the case:
$$sp_{FP} \geq sp_{PBP} \sim sp_{GCP} \geq sp_{FCP}$$

\subsection{Summary}

\begin{figure*}
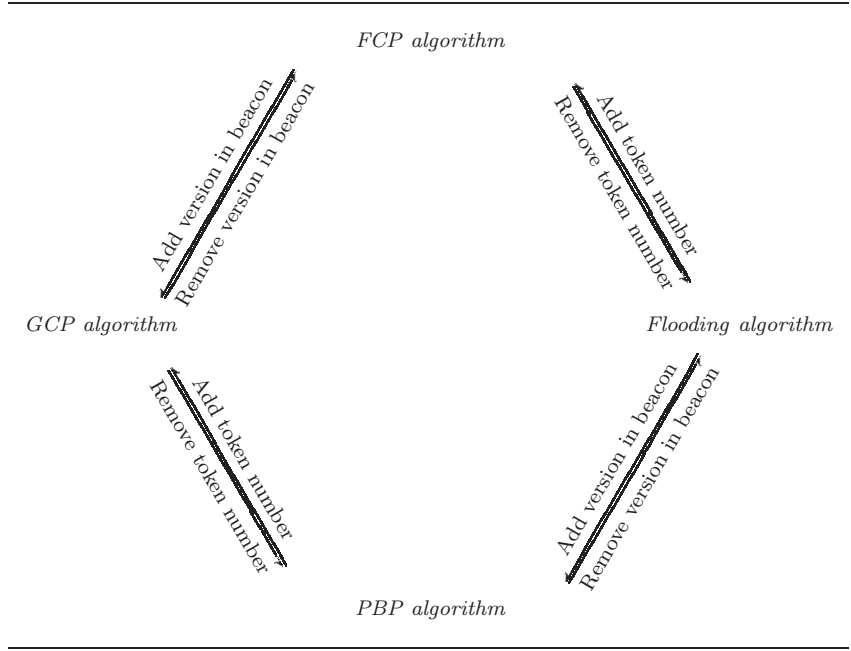

{\footnotesize\begin{center}
\begin{tabular}{ccccc}
\hline
\\
 &  & \it{FCP algorithm} &  &  \\
 & \hspace*{-1cm}\begin{minipage}{2.5cm}\begin{center}\rotatebox{60}{$\autoleftrightharpoons{Add version in beacon}{Remove version in beacon}$}\end{center}\end{minipage} & 
  & \begin{minipage}{2.2cm}\begin{center}\rotatebox{-60}{$\autoleftrightharpoons{Add token number}{Remove token number}$}\end{center}\end{minipage} &  \\
\it{GCP algorithm} &  &  &  & \hspace*{-1.5cm}\it{Flooding algorithm} \\
 & \hspace*{-1cm}\begin{minipage}{2.2cm}\begin{center}\rotatebox{-60}{$\autoleftrightharpoons{Add token number}{Remove token number}$}\end{center}\end{minipage} & 
  & \begin{minipage}{2.5cm}\begin{center}\rotatebox{60}{$\autoleftrightharpoons{Add version in beacon}{Remove version in beacon}$}\end{center}\end{minipage} &  \\
 &  & \it{PBP algorithm} &  &  \\
 \\
\hline
\end{tabular}\end{center}}
\caption{Algorithms dependence}
\label{fig:relation}
\end{figure*}

Figure~\ref{fig:relation} presents the relationship between each of the
previous proposed algorithm. It summarizes how to transform one
algorithm to obtain another one.

Table~\ref{table:compar}  summarizes the  theoretical characteristics of the studied protocols.
The additional local information consists in the token value included
in GCP and FCP, and the beacon one is the version number included in
GCP and PBP. The two latest columns are some assumptions according to load
balancing and propagation speed efficiency.

\begin{table*}
{\footnotesize\begin{center}
\begin{tabular}{|l|c|c||c|c|}
\hline
 & {\bf No local } & {\bf No beacon } & {\bf Load } & {\bf Propagation }\\
 & {\bf  additional } & {\bf  additional } & {\bf  balancing} & {\bf  speed}\\
 & {\bf  information} & {\bf  information} & {\bf } & {\bf }\\
\hline
{\it Flooding Algorithm}                        & + & + & -- -- & + + \\ 
\hline
{\it FCP Algorithm}                                      & -- & + & + & --\\
\hline
{\it PBP Algorithm}                                  & + & -- & -- & + +\\
\hline
{\it GCP Algorithm}                             & -- & -- & + + & + +\\
\hline
\end{tabular}\end{center}}
\caption{Algorithms comparative statement}
\label{table:compar}
\end{table*}

Brown and Sreenan~\cite{Brown2006A-New-Model} proposed a model
to compare software update algorithms in WSNs  
According to this model, GCP provides the following
capabilities: 
\begin{description}
\item[Propagation capabilities]
\item[{\it Advertise}] (of the existence of a new version) is provided
  by  the piggy-backing of the version  number in each beacon message;
\item[{\it Transfer/send}] actions are  provided  by the MAC layer  of
  the node;
\item[{\it Listen}] is provided by receiving beacon message with their
  additional information;
\item[{\it Decide}] is provided by the comparison of version number.
\end{description}
\begin{description}
\item[Activation capabilities]
\item[{\it  Verify}] is provided  by   sending a MD5 signature  before
  sending the new software  version  (if the  software appears to   be
  corrupted,  the node resends  a  beacon to request another  software
  copy);
\item[{\it Transfer/send}] is provided by the MAC layer of the node;
\end{description}
As most of this kind of algorithm, GCP does not provide generation
capabilities and high-level activation capabilities. It is possible to
put GCP in the Deluge~\cite{Hui2004The-dynamic-beh} equivalence class
as it is providing the same capabilities according to the previous
cite model~\cite{Brown2006A-New-Model}.

\begin{figure*}
\begin{tabular}{cc}
\begin{minipage}{0.45\textwidth}
\begin{center}
\includegraphics[width=0.8\textwidth]{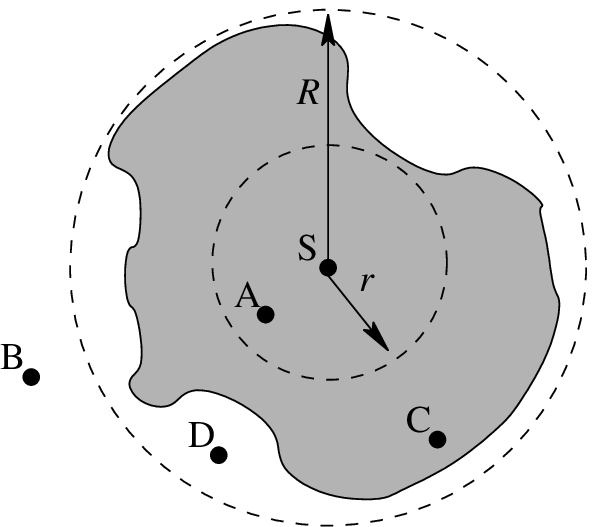}
\end{center}
\caption{Several cases of transmission range.}
\label{fig:rangeDraw}
\end{minipage}
&\begin{minipage}{0.45\textwidth}
\includegraphics[width=0.95\textwidth]{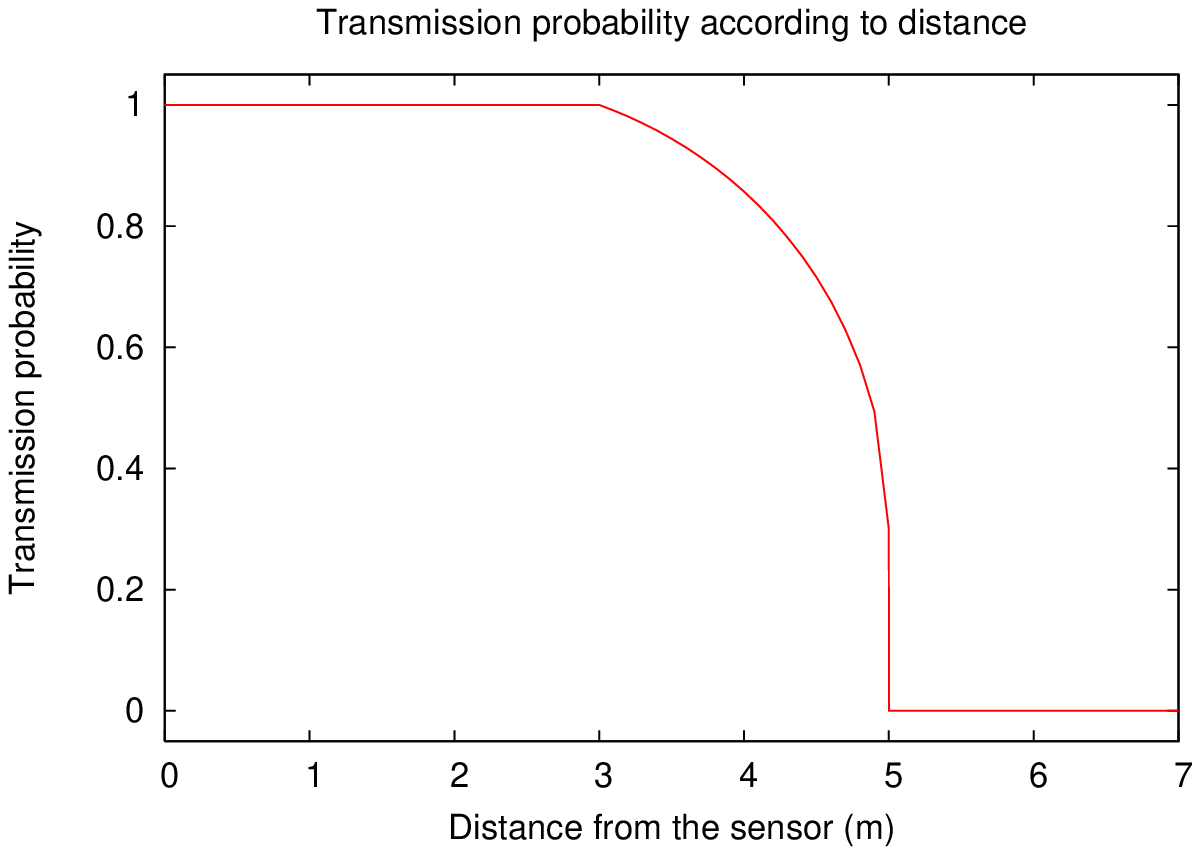}
\caption{Sensor transmission probability according to $r = 3$, $R = 5$, and $P_{min} = 0.3$.}
\label{fig:transmProba}
\end{minipage}
\end{tabular}
\end{figure*}

\section{Simulation results}
\label{toc:simu}

\subsection{System model}

We assume that nodes can communicate only by 1-hop broadcast with nodes
in their  transmission range,
with no collision. For a node $S$, we distinguish two ranges of
transmission,$r$ and $R$, where $0 < r < R$ and $\mathcal{S}_S(r)
\subset \mathcal{S}_S(R)$\footnote{$\mathcal{S}_S(x)$ is the sphere
notation with center $S$ and radius $x$.} ({\it cf.}
Figure~\ref{fig:rangeDraw}). $r$ represents the radius where the
transmission range is uniform, and thus messages sent by nodes
separated by less than $r$ are always received (Node $A$ in
Figure~\ref{fig:rangeDraw}). The second range $R$ represents the radius
where transmission range may be not uniform. No nodes separated by more
than $R$ can receive each other transmission (Node $B$ in
Figure~\ref{fig:rangeDraw}). Thus, nodes separated by a distance
between $r$ and $R$ may or not receive each other transmitted messages
according to the Equation~\ref{equ:range} (Node $C$ and $D$ in
Figure~\ref{fig:rangeDraw}, where the filled shape represented the
transmission area of $S$). In Equation~\ref{equ:range}, $P_{min}$ is
the transmission's lower bound probability parameter for two nodes
separated by $R$. We consider that sensor nodes have equal
communication ranges.  Nodes have a transmission probability defined
as follows, plot in Figure~\ref{fig:transmProba}:
\begin{equation}
\left\{\begin{array}{ll}
1 & \textrm{ if }d < r\\
P_{min} - \sqrt{\frac{R-d}{R-r}}\cdot\left(\frac{R-d}{R-r} -5\right)\cdot\frac{1-P_{min}}{4} & \textrm{ if }r < d < R\\
0 & \textrm{ if }d > R\\
\end{array}\right.
\label{equ:range}
\end{equation}

In our model, we consider mobile sensors. This mobility can
be treated among different mobility models. In order to compare our
results with other ones in the literature, we choose, for the
synthetic workload, the widely used  random way point mobility
model~\cite{Johnson1996Dynamic-source-}.

\subsection{Simulation setup}

\paragraph{Simulator}
In order to evaluate GCP, we developed SeNSim, a software
implemented for mobile wireless sensor-based applications'
simulation. SeNSim is a Java software which allows the creation of
mobile wireless sensor networks and analyses information dissemination
under different mobility and failures scenarios. The simulator also
allows the evaluation of the characteristics related to this protocol
under different mobility, failures, and stimulus scenarios.

In
order to simulate large scale sensor networks  scenarios during a long
period of time,  the designed  simulator is  based  on a discrete-event
system. This software  is   composed  of  two different   parts:   (1)
generation of synthetic workloads and (2) mobile WSN simulation.

\begin{figure*}
\begin{center}
\begin{tabular}{cc}
\hspace*{3mm}\includegraphics[width=0.40\textwidth]{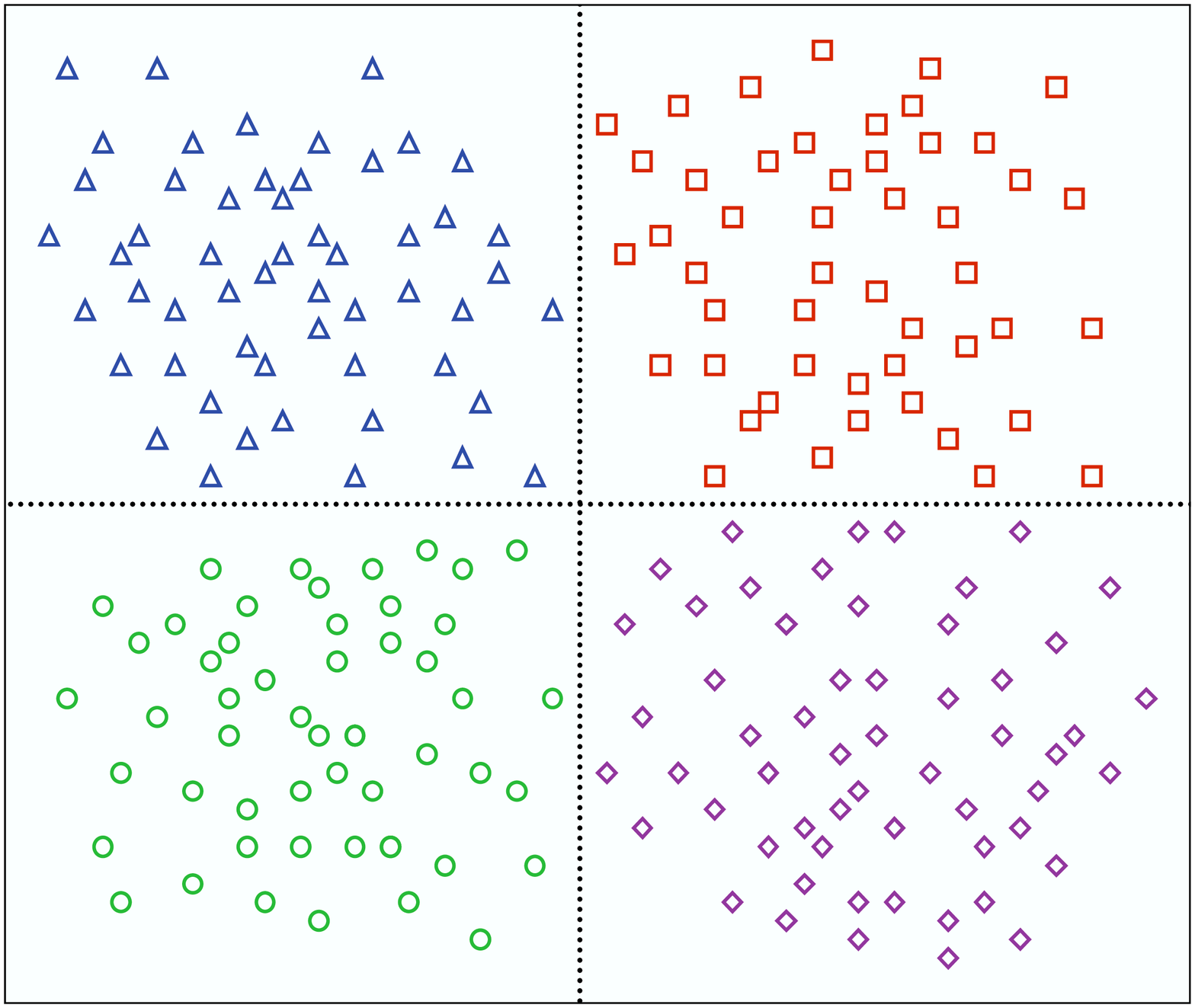}\hspace*{3mm} & 
\hspace*{3mm}\includegraphics[width=0.40\textwidth]{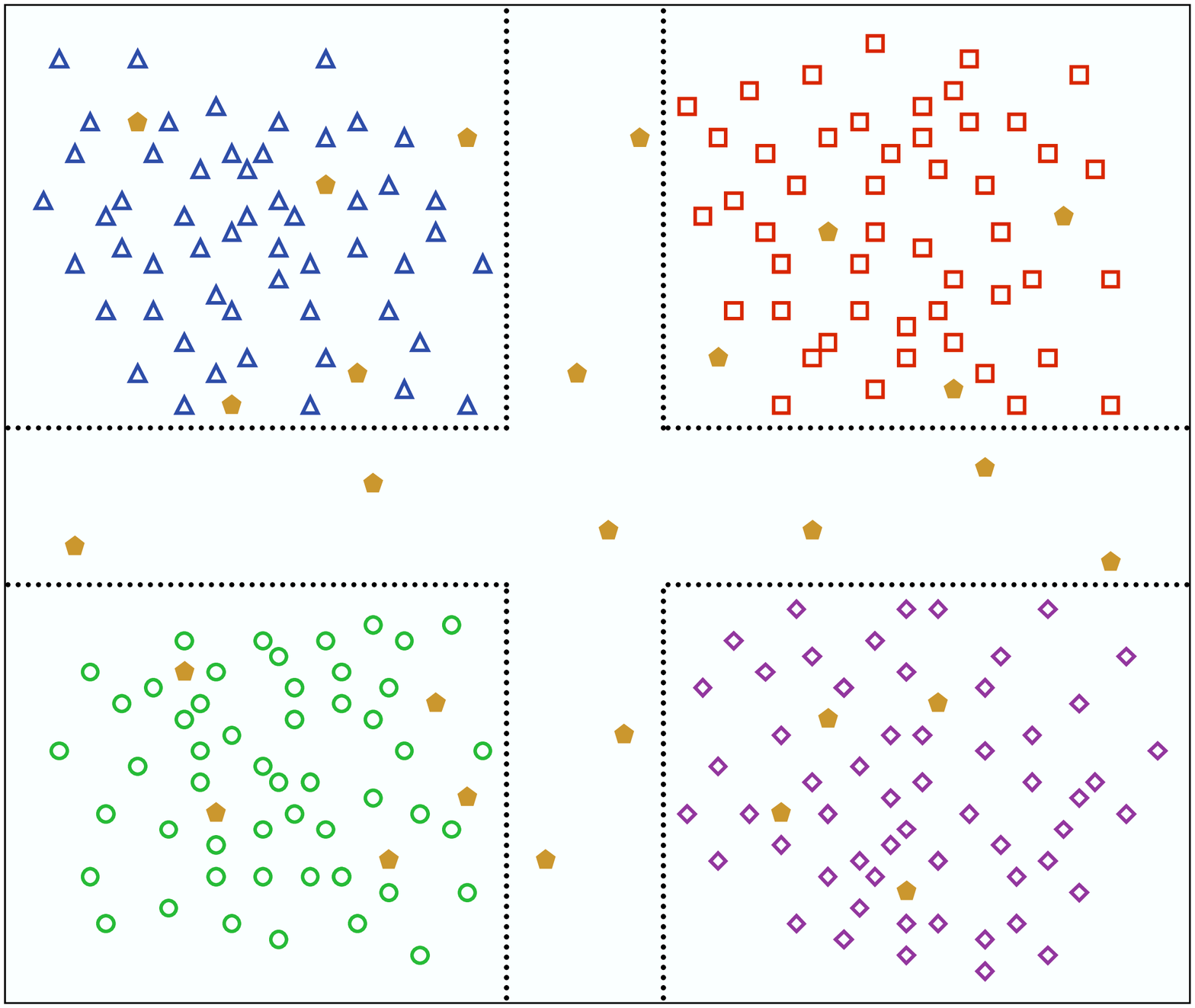}\hspace*{3mm} \\
&\\
(a) Clustering scenario & (b) Socialized clustering scenario 
\end{tabular}
\end{center}
\caption{Examples of synthetic scenarios with four clusters}
\label{fig:scenario}
\end{figure*}

\paragraph{Workloads}
We evaluate different scenarios with the same set
of workloads for comparison purposes. By running different algorithms
on a same persistent trace, we obtain a fair comparison between each
solution introduced in Section~\ref{toc:algo}.

We explore the performance of GCP in various
scenarios. Such scenarios exhibit different clustering
and mobility patterns. Eight synthetic and one realistic scenarios have been
simulated:
\paragraph{Synthetic}
\begin{itemize}
\item {\it  1 cluster scenario --}  One large cluster  composed of 2,000
  sensor nodes in a $250 m \times 250 m$ area.
\item {\it 1 sparse  cluster scenario --}  The large big cluster in  a
  wide area ($1,100 m \times 1,100 m$).
\item {\it 2  cluster scenario  --} Two 1,000  nodes' clusters  in two
  $800 m \times 800 m$ area with a $100 m \times 100 m$ intersection
\item {\it 2 socializing cluster scenario --} Two separated 950 nodes'
  clusters in two $800 m \times 800 m$ area.  These two clusters can't
  communicate   with each other. We   put 100 {\it transmitters} which
  moved in the whole area  ($2,000 m  \times  2,000 m$) to assure  the
  connectivity.
\item  {\it 4 cluster scenario --}  Four separated 500 nodes' clusters
  in four $550 m  \times 550 m$ areas. Connectivity  is assured by the
  common  border.        These    areas    are    set       as      in
  Figure~\ref{fig:scenario}(a).
\item  {\it  4 socializing cluster  scenario  --}  Four  separated 475
  nodes' clusters in   four $550 m   \times 550 m$ areas.  These  four
  clusters  can't  communicate  with   each  other. We  put  100  {\it
    transmitters} which moved in the whole area ($1,300 m \times 1,300
  m$) to  assure  the   connectivity. These   areas  are set    as  in
  Figure~\ref{fig:scenario}(b), where  transmitters are represented by
  a filled pentagon.
\item {\it  9 cluster scenario --}  Nine separated 250 nodes' clusters
  in four $400 m  \times 400 m$ areas.  Connectivity is assured by the
  common border.
\item  {\it 9 socializing   cluster  scenario--}  Nine  separated  240
  no\-des'  clusters in four  $400 m  \times 400  m$  area. These nine
  clusters can't   communicate with   each other.   We put   90   {\it
    transmitters} which moved in the whole area ($1,500 m \times 1,500
  m$) to assure the connectivity.
\end{itemize}
\paragraph{Realistic}
\begin{itemize}
\item   {\it MIT  Campus   scenario --}   In   order to  evaluate  the
  performance of GCP with a realistic  movement behaviour, we used the
  mit/reality data  set~\cite{crawdad}  from  CRAWDAD. This  data  set
  provides captured   communication,  proximity, location  and activity
  information  from    100 subjects  at MIT  over   the course  of the
  2004-2005 academic year.
\end{itemize}

We have used the following parameters:
\begin{itemize}
\item The time is discretized by millisecond.
\item Each sensor node:
\begin{itemize}
\item is initially randomly placed inside its defined area for
synthetic workloads.
\item sends  a    beacon  periodically every 100   ms   (common period
  encountered in the literature).
\item has the transmission ranges  set as follows:  $r = 3 m$  and $R =  5 m$ and  a
  minimum transmission probability inside $R$, $P_{min} = 0.3$.
\item is  mobile, following  a {\it   Random  Way Point} strategy  for
  synthetic workloads, with  a  maximum pause time  of $100  ms$, each
  movement duration between $100$ and  $500 ms$, with a speed included
  between  $0.8 m.s^{-1}$ ($2.88  km.h^{-1}$)  and $2 m.s^{-1}$  ($7.2
  km.h^{-1}$) (equivalent to  human walking speed). Every movement  is
  bounded by the  defined area. The border rules  are  defined as each
  node bounce back according to the bisector of incidence angle.
\item has a number of tokens, for each  code propagation, according to
  simulation configuration describes below.
\end{itemize}
\item Simulation last around $50,000 ms$.
\item A new version is sent  to a sensor picked  up at random after $1
  s$ of simulation.
\end{itemize}

In order to compare the efficiency of the four algorithms, we compared
them along the following metric, previously introduced
Section~\ref{toc:theory}:
\begin{description}
\item[Code propagation speed] We observe the number of no\-des owning
the newest version of the software during all the simulation, and plot
these values in Section~\ref{toc:cs} for each scenario and algorithms.
\item[Load balancing] At the simulation termination, we extract from
each node the number of times it sends the software. Results are
depicted in Section~\ref{toc:lb} for each scenario and algorithms.
\end{description}

\subsection{Convergence speed}
\label{toc:cs}

For each simulation scenario, we have followed the propagation
advancement, and extracted at each time, the number of sensors owning
the newest version of the software.

\begin{figure*}
\begin{center}
\includegraphics[height=0.3\textheight]{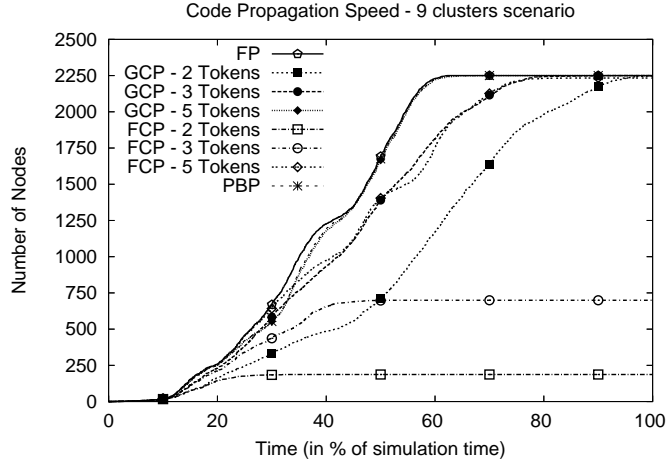} 
\end{center}
\caption{Convergence speed for the largest clustered scenario}
\label{fig:cps9c}
\end{figure*}

\begin{figure*}
\begin{center}
\includegraphics[height=0.3\textheight]{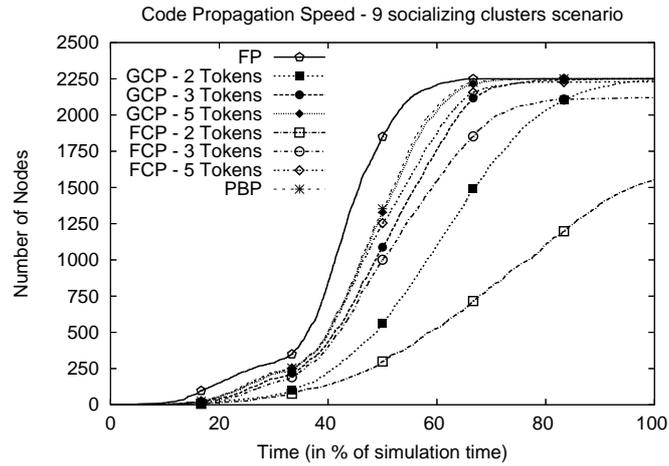} 
\end{center}
\caption{Convergence speed for the largest socialized scenario}
\label{fig:cps9s}
\end{figure*}

\begin{figure*}
\begin{center}
\includegraphics[height=0.3\textheight]{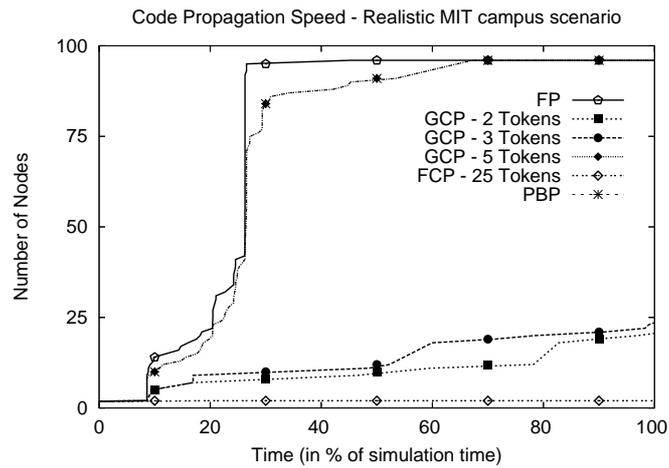} 
\end{center}
\caption{Convergence speed for realistic scenario}
\label{fig:cpsreal}
\end{figure*}

Figures~\ref{fig:cps9c},~\ref{fig:cps9s}~and~\ref{fig:cpsreal} present the results according to time, for three scenarios ordered as above:
\begin{enumerate} 
\item 9 clusters scenario organized in the same way than in Figure~\ref{fig:scenario}(a);
\item 9 socializing clusters scenario organized in the same way than in Figure~\ref{fig:scenario}(b);  
\item the MIT campus realistic workload scenario.
\end{enumerate}
For each one, we have plotted the code propagation speed for GCP,
FP, PBP and FCP. Each
synthetic scenario has approximately the same propagation behaviour as the 9
socializing clusters scenario (cf. Figure~\ref{fig:cps9s}). Due to the
space constrain, we have represented the results for only two synthetic scenarios. The  9
clusters scenario is presented here to illustrate the fact that GCP outperforms
the FCP algorithm.

In the flooding algorithm, each time a sensor node meets another one, it
sends its own software. As explained before, the flooding code
propagation speed can be considered as the ideal transmission speed
and is taken as reference thereafter. For the two algorithms with
token rules (Forwarding control mechanism), we have plotted the results obtained by using $k$ tokens
a node, where $k$ is respectively equal to 2, 3 and 5. We do not
represent results for more than 5 tokens as in the case majority, GCP
tends to approach the ideal reference by using only 5 tokens (Each
synthetic simulation system counts around 2,000 sensors, so for all $k
\sim \log(2000) = 3.3$~\cite{KerMasGan03IEEEtpds}).

Regardless of the number of tokens chosen, in each scenario, GCP outperforms
FCP as far as propagation speed is concerned. Taken flooding
algorithm as ideal reference, Figure~\ref{fig:cps9c} shows different
inflection points. These are due to the software transmission from a
cluster to another. This is clearly denoted in Figure~\ref{fig:cps9c}:
when almost half of sensors nodes have picked up the software's newest
version, there is a period during which the newest version are moving
from a cluster to the other.

Figure~\ref{fig:cpsreal} presents the propagation speed during the
realistic scenario. It is interesting to observe that FCP algorithm is
not efficient in this scenario. In real life, some people are together
most of the time.  As nodes used FCP algorithm, they are not aware from the
remote nodes' version, and may spend all their tokens for the same
node. GCP with a small number of tokens (2 and 3 here for instance) remains better than FCP with a large number of tokens. With only 5 tokens
per node, GCP is almost as efficient as the flooding algorithm with respect to 
propagation speed.

For each scenario, PBP is significantly slower than the flooding one and is
always equivalent to GCP with 5 tokens per node.

Simulation results show the propagation speed efficiency of GCP
according to the FCP algorithm for the same number of tokens and to the
 flooding algorithm as ideal reference. We have measured as well
the network load balancing, presented in the next subsection.

\subsection{Load Balancing}
\label{toc:lb}

\begin{figure*}
\begin{center}
\includegraphics[width=0.7\textwidth]{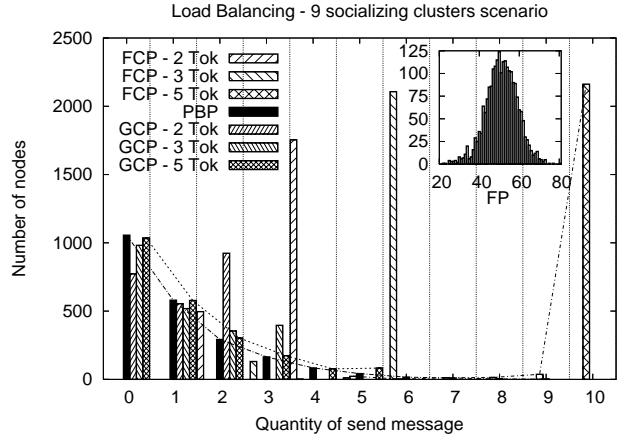}
\end{center}
\caption{Load balancing for the largest socialized scenario}
\label{fig:lb}
\end{figure*}

In order to estimate the benefit of GCP in term of load balancing, we
have collected for each simulation (scenarios and algorithms), the
number of sent software messages by node.  As each simulation have the
same network load behaviour, Figure~\ref{fig:lb} presents the results
for the 9 socializing clusters scenario. For each number of message
sent (represents by the X-axes), we have represent for all simulations
the number of nodes which have sent exactly this number of time its
software. As the flooding consumes much more message than the three
other algorithms (PBP, FCP and GCP), the network load with flooding is
represent in the upper-right corner of Figure~\ref{fig:lb}.

 For each scenario, the benefit of GCP or FCP compared to the flooding
 algorithm is clear. When considering the number of software binary sent, the two other algorithms save
 between 82 \% and 93 \% of messages for FCP and more than 98 \% for
 GCP for a 50 seconds simulation only. As we have presented above, the
 number of software sent message increase according linearly to time
 with the flooding algorithm.

By focusing only on the three other algorithms, main plots of
Figure~\ref{fig:lb} show the benefits of GCP according to PBP and
FCP. In fact, as FCP is not aware of the remote node's version, the
local node sends its own version as long as it still possesses
tokens. That implies the number of software sending messages with FCP
algorithm is almost constant, corresponding of two times the number of
tokens owns by each node ($k$ tokens for the first version plus $k$
tokens for the newer version: in these simulations, the software is
updated only once).

With GCP and PBP, the current local version number is sent in the
beacon message (Piggy-backing mechanism). So, nodes will not send the first version. They only
send the newest version, and only if the beacon sender does not owned
the latest version. The total number of software send messages in the
network is almost equivalent to the number of participating
nodes. Moreover, as every node in the transmission range of a sending
node receive the sending version, freely for the sender, the network
load benefits using GCP is decrease all the more. Comparing GCP with 5
tokens a node and PBP, which have the same propagation speed, we
observe that near 5~\% of the nodes have send the software more
that 5 times, contrary to GCP where nodes have consumed at most 5
tokens.

We do not represent the load extracted from simulation on the realistic
trace because it shares the same aspect as the synthetic ones.

\section{Related works}
\label{toc:relatedWorks}

This section presents various works in two related axes namely: data
dissemination (or broadcasting) and software update (or reprogramming
/ code propagation). Among those, we focus on the approaches relying
on gossip-based algorithms.

\subsection{Dissemination in WSN} 
Reliable data dissemination shares some goals and assumptions with
code propagation. Reliable broadcasting consist in a one-to-all
message dissemination (one entry point in the system sends some
information to all participating nodes in the network).

Vollset and al.~\cite{Vollset2003A-Survey-of-Rel} propose a
classification of reliable broadcast in two classes: deterministic and
probabilistic protocols. Deterministic protocols attempt to enforce
hard reliability guarantees, as probabilistic protocols provide
guaranteed delivery with a certain probability. This paper concludes
that deterministic approach tends to have bad tradeoffs necessary
between reliability and scalability/mobility, instead of probabilistic
protocols, which do not provide deterministic delivery guarantees. Due
to the constraints present in Section~\ref{toc:intro}, we focus on
probabilistic protocols to obtain a reliable efficiency in scalability
and mobility models.

 Some study used geographical information to increase broadcast
 efficiency (load balancing, propagation speed,
 \ldots). In~\cite{Subramanian2006Broadcasting-in}, several degrees of
 local information has introduced in broadcast protocols. Subramanian
 and al. present three degrees of knowledge: no geographic or state
 information, coarse geographic information about the origin of the
 broadcast and no geographic information, but remember previously
 received messages. Authors conclude that local information have a not
 negligible role in broadcasting in WSN. As our contribution, several
 works tends to increase the broadcast efficiency without using
 locality.~\cite{Intanagonwiwat2003Directed-diffus,
 Li2004BLMST:-A-Scalab, Mitton2005Broadcasting-in}

Another probabilistic broadcast is based on the gossip-based
model~\cite{Kyasanur2006Smart-Gossip:-A,
Wang2005Reliable-Gossip}. Gossip-based protocol is a promising model
for WSN as presented in Section~\ref{toc:intro}. Gossip as a general
technique has been used to solve several problems (data management,
failure detection, \ldots). In the end of this subsection, we present
different gossip-based dissemination protocols for WSNs.

In~\cite{Kyasanur2006Smart-Gossip:-A}, a reliable broadcast service
based on the gossip model is presented. Using probabilistic
forwarding, gossip is used to self-adapt the probability of
information send for any topology. But, in most case, mobility is not
taken into account.

Wang and al.~\cite{Wang2005Reliable-Gossip} proposed a reliable
broadcast protocol for mobile wireless sensor network. By using
clustering technique and gossiping, this protocol has high delivery
ratio and low end-to-end delay.  Another noteworthy work is a reliable
multicast protocol for mobile ad-hoc network (MANET), called {\it
Anonymous gossip}~\cite{Chandra2001Anonymous-Gossi}. Based on reliable
gossip-based multicast for wired networks, gossip is used to obtain
some information about message which node has not yet received. In
these two last references, authors are studying MANETs, which have more
resources than sensor nodes.

Our works is a transversal axis of these last studies. We are taking
into account mobility and power consumption of WSN by using
gossip-based paradigm.

\subsection{Code propagation in WSN}

Recently, gossip-based model has been used in code propagation
service~\cite{Levis2004Trickle:-A-Self,
Wang2006Gappa:-Gossip-B}. Trickle~\cite{Levis2004Trickle:-A-Self} is a
code propagation algorithm using a "polite gossip" policy. This
algorithm permits to propagate and maintain code updates in WSN. It
regulates the transmission by gossiping code meta-data. If the
information is up-to-date, the receivers stay quiet. Otherwise, if the
information is issued from an older version, the gossiper can be
brought up to date, and similarly.

Other works propose several protocols to apply code
propagation. Deluge~\cite{Hui2004The-dynamic-beh} assure to
reprogramming the network by a reliable data dissemination
protocol. Authors argue that Deluge can characterize its overall
performance. Notably, they assume that it may be difficult to
significantly improve the transmission rate obtained by
Deluge. Kulkarni and Wang proposed
MNP~\cite{Kulkarni2004MNP:-multihop-n}, a multi-hop reprogramming
service, by splitting code into several segments. Using pipelining and
sleep cycle, this algorithm also guarantees that, in a neighbourhood,
there is at most one source transmitting the program at a
time. Recently, these authors present
Gappa~\cite{Wang2006Gappa:-Gossip-B}, an extension of MNP. Using an
Unmanned Ariel Vehicle (UAV), this algorithm can communicate parts of
the code to a subset of sensor nodes on a multiple channel at
once. The protocol ensures that at any time, there is at most one
sensor transmitting on a given frequency.

In the best of our knowledge, none of work treats about code
propagation in \emph{ mobile} WSN.

\section{Conclusion}
\label{toc:conclu}

In this paper, we have proposed a software update algorithm for mobile
wireless sensor networks. As the best of our knowledge, tackle code
propagation in mobile WSN has not been done before. Based on leverage
works on epidemic protocols and on similarities and differences
between P2P systems and mobile WSNs, {\it Gossip-based Code
Propagation} algorithm tends to outperforms classical dissemination
algorithms, with only a small overhead by adding little extra
information on sensor nodes and in beacon messages.

We have exposed the benefit of GCP on several simulation scenarios,
compared as three other dissemination algorithms: one ideal in speed
convergence but with a large number of software send messages and,
therefore, very high power consumption, another one based on
forwarding control and a last one based on piggy-backing message.

For each of these algorithms, GCP obtains an important profit
accordingly to the little overhead information. With a clearly load
balance through the network, GCP can disseminate the new software with
almost the same propagation speed than the ideal one.

One of these work perspectives consists to include the {\it Delay
Tolerant Networks} (DTN) paradigm into GCP in order to take into
account and optimize the free receptions of the software due to the
omnidirectionnal wireless transmission.

\bibliographystyle{abbrv}
\addcontentsline{toc}{section}{References}
\bibliography{gcp}

\newpage

\tableofcontents

\end{document}